# Green Telecom Metrics in Perspective

Daniel Kharitonov
Juniper Networks Inc
Sunnyvale CA, USA
dkh @ juniper.net

*Abstract*—A fast and parallel evolution of ways to measure and assess energy efficiency in telecom has resulted in an entangled web of drafts and recommendations originating from government, research, and standards organizations. This paper focuses primarily on so-called "large network equipment" metrics and intends to capture state-of-the-art in this area of green communications. Competing approaches towards efficiency assessment are studied for their applicability and completeness, with special emphasis on topics relevant to future subject studies.

## I. INTRODUCTION

At this particular moment in human history, little doubt remains that we need to focus on sustainable development in all aspects of technology, including information and communications infrastructure. This is why rapid growth of the telecom energy footprint [1] is causing major concern from public, government, and research organizations, and warrants close attention to "green communications" promised by smarter and leaner designs. However, the laws of economy dictate that such material improvements should also be supported by market requirements. Such requirements mean a transition from intangible (marketing-based) to tangible (measurable) practices in rating and selecting telecom devices on energy efficiency [2]. It is no coincidence that large service providers and vendors are among the first apologists of this transition—the former are driven by the need to stop proliferation of electric bills already measured in billions of kilowatt-hours [3], while the latter are fascinated by exploring synergy between energy efficiency and product performance. Therefore, the new agenda of green telecom is no longer "how to evaluate sustainability" but "what efficiency metrics to use."

In the following paper, we look deeper into the process of creating telecom energy efficiency metrics and note successes and pitfalls observed over the last years. We also intend to classify and categorize such metrics, discussing the pros and cons of different approaches. Finally, we pay special attention to uncharted territories and areas for future research.

## II. DOES "GREEN" EQUATE TO "LOW ENERGY"?

In theory, a study subject of "green telecommunications" should be all-encompassing and range from managing hazardous substances to lifecycle carbon assessment (LCA), of which energy consumption is merely a runtime component. In practice, vendors cannot ship equipment that does not conform to major compliance directives such as the European Union's Restriction of Hazardous Substances (RoHS) or Waste Electrical and Electronic Equipment (WEEE) directives. Therefore, eco-friendly materials and disposal practices are subject to sporadic activity in research and development as new requirements emerge. This activity stops shortly after the minimum compliance level is reached.

On the opposite end of the spectrum, lifecycle analysis (LCA) and carbon footprint minimization (as described in ISO 14000 series documents [4]) call for ongoing and steady improvements. However, if we focus on full carbon profiling, consistent content tracking from cradle to grave can become very complex for simple products like wood and paper [5], and even more so for telecom devices with parts produced by a deeply nested stack of international suppliers.

This is why the discussion around "green telecom" and "green metrics" for information communications technology (ICT) products often revolves around runtime energy consumption and carbon generation. In this paper, we will adopt a similar stance by deemphasizing the distinction between sustainability and energy use. Instead, we will assume the "green telecom device" to be the one offering the highest energy efficiency.

## III. GENERAL METRIC DESIGN

As our goal is ultimately to grade telecom systems according to energy use[1], we need to establish measurable efficiency properties otherwise known as metrics. The definition of efficiency assumes doing more work for less energy, so a reasonably good metric should rate functionality against energy consumption. The idea behind establishing metrics is very simple: a product with "greener" credentials should have more chances in the marketplace, which in turn should encourage other vendors to innovate in this particular area of human knowledge; less efficient products are to be depressed and leave the scene (Figure 1). The need for metrics stems from the fact that oftentimes efficiency ratings are not evident from the product's look and feel, and it helps to educate customers on making energy-aware purchasing decisions.

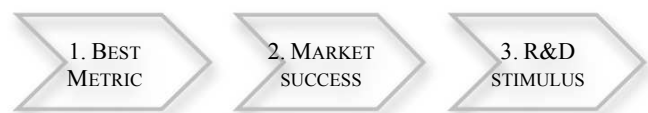

Figure 1.  Desirable product lifecycle flow when using metrics

---

[1] While it is also possible to define metrics for networks as a whole, this study focuses on individual systems

But before we get down to metric design, we need to define the notion of *work,* or rather *useful work*. Surprisingly, the definition of efficiency for telecom devices is a hard problem to solve across the wide variety of information communications technology (ICT) systems, and it still remains open for many device types. The most obvious issue is that most telecom systems are not designed to transform energy, i.e., the energy output of most telecom devices (in the form of electric current, radio waves, and photonic beams sent into connected systems) can be small relative to energy consumption.

Instead of producing energy or mechanical work, telecom systems predominantly provide information services such as transport, security, and signal conversion. This is precisely why pure energy transformation metrics (such as Green Grid's PuE [6] and EPA power supply ratings [7]) are not appropriate for telecom devices. Indeed, while every ICT system uses internal or external energy conversion parts, they alone cannot determine the final efficiency of a telecom product. A system with good power supplies can be very bad at processing data, and this will be hidden from the consumer looking only at the power supply or cooling efficiency label.

It is also easy to see that if a telecom product operates across multiple modalities, a multi-dimensional utility metric might still not be a good way to compare product A to product B. For example, let's consider a home office gateway. Such a device may combine a broadband modem, wireless access point, Ethernet switch, and security subsystem. This gives at least four distinct areas of "useful work" which need to be captured in the metric. However, a four-dimensional comparison between similar products is difficult (if not confusing), and can be further aggravated by the variability of competitive designs featuring diverse combinations of functions, ports, and speeds.

Luckily, the ICT industry has known about this problem for a long time and has a ready solution. Instead of being labeled with a straight efficiency metric similar to US EPA mile per gallon rating for cars [8], a telecom product (or rather, a class of products) can be given a series of "not to exceed" allowances designed to capture specific dimensions and functions. For example, the European Committee Code of Conduct for Broadband Equipment (EU CoC) defines such allowances for every WAN, LAN, and auxiliary network interface [9]. A system with an arbitrary combination of those interfaces gets its energy consumption ceiling from an arithmetical sum of allowances, and the EU CoC program intends to update tiers to keep pace with technology.

Indeed, considering the ongoing progress in silicon scaling and energy consumption (otherwise known as Dennard's scaling law [10]), allowances cannot be left static and should be updated frequently via the high-touch process of reaching consensus between vendors, industry experts, and government organizations. This process should be repeated periodically in order to put obsolete technology to disadvantage.

However, despite wide availability and apparent extensibility (new cells can be easily formulated for emerging functionality types), allowance-based energy efficiency schemes present one major problem—they are designed not to encourage winners but to discourage losers, which raises questions about their effectiveness.

Although one can technically describe an allowance system (such as the EU CoC program) as a "metric" because it measures the maximum energy budget related to a specific telecom system's tier, it is merely a ceiling that is formulated to fit the vast majority of shipping systems. Since it is common to have product cycles overlapping between vendors, cut-off allowances are normally defined as lowest common denominators between several generations of hardware in active use, and thus fail to foster innovation (Figure 2).

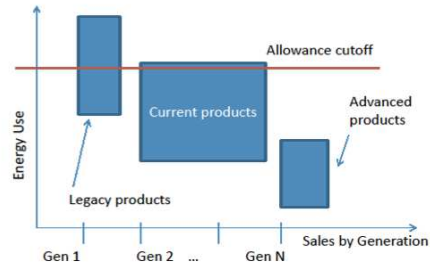

*Figure 2. Sales of product by generation relative to cut-off*

In other words, if someone designs a massively effective broadband gateway that is 10x more efficient than the competition, it would be classified into the same "pass" bucket as 98 (or even a higher) percent of systems on the market.

In order to get around this issue, an allowance program can be complemented with program grades (gold, silver, bronze, etc.), but this would only partially solve the problem, as the target metric would remain discrete. Depending on the state of the industry, it would be quite possible for grades to lump products with different efficiencies together, or move products with a small difference in efficiency into adjacent bins. When looking at "golden" versus "silver" classifications, customers might have little idea about what this really means to their network and why it is important.

Thus, we can conclude that an allowance system is generally suboptimal and should be used as an effort of last resort when productivity-based criteria cannot be established.

Whenever possible, telecom systems should be gauged by single-dimensional analog metrics that reflect the productivity ("useful work") of a product. A metric of this type does not require periodic updates to catch up with technology; instead, it allows every organic improvement in device technology to be immediately visible in a datasheet. Armed with this information, the user can make an educated decision on whether the product has a critical advantage over the competition or not.

Early in the standardization process, it became quite clear that at least one large category of telecom devices (namely performance-oriented transport systems) was relatively easy to describe with respect to "useful work" and hence could be judged using one-dimensional continuous metrics.

This category includes time-division multiplexing (TDM) equipment and carrier-class routers, switches, and security devices that form the backbone of today's Internet and intranets. Such devices form a convenient target group because they have one clear modality—effective capacity. Since cumulative egress output of their interfaces and runtime energy consumption can be related to each other, this forms the natural basis for analog metric design.

Throughout this paper, we will further consider only the analog metrics, with a focus on their design and robustness.

## IV. PEAK METRICS

Peak efficiency metrics are typically well understood by the general public, with plenty of examples known from everyday life. Whenever an automaker announces a new car or an aerospace consortium launches a new airplane, chances are we will hear some peak efficiency figure expressed in terms of fuel consumption per distance or carbon emissions per passenger-kilometer, measured under the most favorable conditions. While we all know that cars are not always driven on highways and airplanes are not always flown fully loaded, such metrics offer technology assessments that are simple, transitive, and repeatable. This means that a car with a better efficiency rating within its class is likely to feature the most advanced construction and fuel management technology, and the airplane with less carbon emissions per seat will statistically be less harmful for the environment across a wide range of flight profiles.

For telecom TDM equipment such as Synchronous Optical Networking (SONET) and Synchronous Digital Hierarchy (SDH) switches as well as newer Optical Transport Network (OTN) platforms, the peak metric can be obtained as simply as a sum of the nominal speed of interfaces divided by runtime energy consumption. This is due to the fact that time slot switches do not have a concept of "performance," just capacity. On the other hand, equipment dealing with multiplexing the fixed length or variable length pieces of information (packets) such as routers and switches may exhibit a vastly different behavior. Under certain conditions, for example, a device may become overloaded with traffic and fail to deliver full throughput even if the sum of input packet streams does not exceed the available egress bandwidth. Such limitations may exist per design (e.g., when optimized to handle traffic only in certain directions) or due to genuine performance limits. Therefore, packet-based equipment requires well-chosen test conditions and methodology to ensure that efficiency tests produce sensible results.

First introduced to the industry in 2008 by the Japanese Energy Efficiency standards committee (METI), the peak metric set and methodology recommendation for efficiency assessment of packet-based routers and L2 switches included methods and calculations to construct metrics based on the system's maximum measured performance weighted for different packet sizes [11].

Being clearly an early and visionary achievement, the METI recommendation had all the traits of a solid metric design and even included some of the measurement methodology. Routers and switches were grouped based on the target functions and performance parameters. The resulting metric was clearly analog and allowed to identify class leaders (also known as "Top runners"). The specification was later extended to include other equipment types as well.

Unfortunately, the METI specification was never widely adopted abroad. This can be attributed to the overly complex and incomplete nature of test methods, as some equipment was specified to be tested at fixed packet sizes, some at variable sizes, and diverse capabilities of small routers and switches were given a set of allowances. At the same time, the specification did not differentiate between variable-load and idle-load power state behavior, which made it difficult to reproduce METI test results outside of Japan.

Later that same year, a team of Lawrence-Berkeley National Labs, Ixia, and Juniper Networks researchers introduced ECR [12]—a simple peak metric defined as a ratio of effective maximum throughput T (measured at egress) to effective energy consumption E at fixed packet size:

$$ECR = E / T \quad (Watts/Gbps) \qquad (1)$$

This formula has a physical meaning of full duplex throughput that can be observed in the field on a fully loaded system under test. Unlike the METI metric, ECR does not give credit for ingress traffic (only full duplex data is accounted for) and does not require separate measurements for varying packet sizes. In the first ECR revisions, packet size was dependent on class.

Possibly the main asset of ECR was a well-developed and formally defined test methodology, suitable for independent testing. This metric was intended to plant reproducible efficiency estimates into datasheets of packet products at the cost of a relatively simple measurement procedure that included only two runs—one to find the effective throughput (non-drop rate) and the another to find average energy consumption during sustained peak load[2].

Per design, any peak metric provides a simple assessment of a product's technology. For telecom devices, improvements in peak efficiency can be related to both silicon[3] and architectural product improvements (see Table 1 for comparison between core network devices).

*Table 1. Efficiency Progress in Core Routers 2002-2012*

|  | T640 | T1600 | T4000 | PTX* |
|---|---|---|---|---|
| Year | 2002 | 2007 | 2011 | 2012 |
| ECR, W/Gbps | 14 | 9.7 | 3.54 | 1.54 |
| * optimized for MPLS label-switching router (LSR) architecture ||||||

Today, peak efficiency metrics are used extensively during product launches and in datasheets, and it is hard to find a new router or switch that is not being described in peak efficiency terms. Standardized test process and methodology ensures that such numbers (when provided) can be easily compared and verified.

---

[2] ECR 1.0 also included a definition of variable load metric; however, this section was practically never used.

[3] Note that network processors (NPUs) have less regular structure relative to general-purpose processors. This forces network routers to trail the efficiency curve of other computing devices.

## V. Variable Load Metrics

While peak efficiency metrics were quickly becoming a de facto standard for new product introductions, this was clearly not enough for large telecom operators. The issue at hand was very simple—in real-world networks, products never operate at full sail. Although modern router/switch designs are optimized to sustain wire speed operation and traffic bursts, average system load on large intervals rarely exceeds thirty percent and often fluctuates around fifteen percent or even less [13]. Taking this fact into account, the typical router/switch elasticity does not look that good (see Table 2), as a lightly loaded system decreases its energy consumption only by a small fraction.

*Table 2. Load-Proportional Energy Response of a Router*

| Load, % | idle | 10 | 30 | 50 | 80 | 100 |
|---|---|---|---|---|---|---|
| Power, watts | 768 | 790 | 801 | 816 | 842 | 863 |

Presumably, proliferation of the variable load metrics designed to emphasize energy elasticity should benefit designs with relevant hardware capabilities (such as IEEE 802.3az [14]). Therefore, load-proportional metrics were expected to dominate the service provider business, where energy costs are a significant part of operating expenses.

And indeed, it was a U.S.-based service provider, Verizon, that delivered the first publicly introduced variable-load metric in 2008 as part of the initial revision of its Network Equipment Building System (NEBS) TPR.9205 specification [15].

This metric (TEEER) featured an unusual design, using a logarithmic scale in the form of *- log (P/T),* where *T* was defined as "forwarding capacity" or "the number of bits per second that a device can be observed to transmit successfully on the correct egress interface," and *P* was assumed to be a weighted sum of energy consumptions measured over three reference points (0, 50, and 100 percent load).

At the same time, the first (and all subsequent) revisions of TPR.9205 did not explicitly define a test procedure for packet equipment. In particular, Verizon's draft did not have definitions and tests for traffic composure, topology, and maximum forwarding capacity. This gap was apparently driven by the desire to use Independent Test Laboratory (ITL) network that was normally charged with all NEBS tests.

In theory, TPR.9205 should work just fine. Although the TEEER metric was not designed to carry any physical meaning and was expressed as an abstract number, it was still an analog efficiency measure, which ensured that a system with better energy elasticity will show a better result. Meanwhile, an ITL could use just about any test methodology and as long as it was the same across all tested systems, the results should remain comparable.

The weak point in this plan was outside of Verizon and within the ITLs themselves. Compliance test labs were well equipped for environmental certifications, but were not outfitted with expensive test equipment and personnel to run performance tests on telecom devices. Therefore, any results they could deliver were superficial at best and might not properly reflect the elasticity of a system under test.

Understanding the importance of sound test methodologies, the North American Alliance for Telecommunications Industry Solutions (ATIS) took the chance to improve the quality of variable load metrics within its own set of TEER efficiency documents.

ATIS routing and switch specification 0600015.03, published in 2009 [16], featured a metric design virtually identical to Verizon's TEEER (minus the logarithm part) and test specifications borrowed from ECR version 2.0. This enabled any interested party to test variable load efficiency with good precision and repeatable outcomes. However, this specification was also not free of issues.

Where Verizon used just one set of weights and load points (0, 50, and 100%) weighted at 0.35, 0.4, and 0.25 respectively, ATIS introduced a matrix of load points and weights for various equipment types, which caused widespread confusion between TEEER and TEER specs.

Adding to the confusion was the fact that ATIS left the division of maximum throughput by weighted energy consumption in place. This meant that if someone treated the ATIS metric as an estimate of efficiency expressed in gigabits per watt, the resulting number would be higher than the maximum theoretically achievable (T/E=1/ECR) figure.

To understand this issue, we need to look closely at how well-formed variable load metrics (like EPA fuel economy) are constructed (Fig 3).

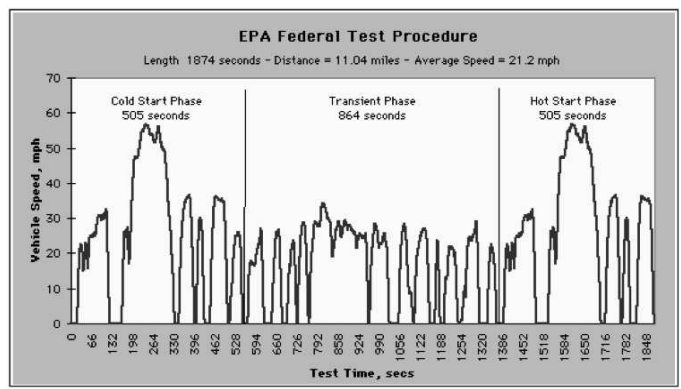

*Figure 3. U.S. EPA vehicle test profile (source: fueleconomy.gov)*

In this graph, a car is subjected to a start-stop speed test, while a metric is based on the cumulative distance covered (in miles) versus cumulative fuel consumption (in gallons) during the test. If EPA had chosen to relate the distance that could have been covered at cruising speed (55 miles per hour) to cumulative fuel consumed in start/stop operation, the results would be completely unrealistic.

A resolution to these issues is expected to come by the end of 2012, when a new revision of ITU-T L-series recommendations and ATIS should converge into a new formula for variable load efficiency and simplified load/weight matrix:

$$\text{EER-VL (Gbps/watt)} = \frac{(\alpha * T_f + \beta * T_r)}{(\alpha * E_{100} + \beta * E_r + \varepsilon * E_i)} \quad (2)$$

where $T_f$ = full throughput, $T_r$ = reduced throughput, $\{E_{100}, E_r, E_i\}$ = energy consumption at full load, reduced load, and idle load, $(\alpha + \beta + \varepsilon = 1)$.

This expression has a physical meaning of calculating the ratio of effective throughput to the effective energy consumption that was recorded during a use cycle consisting of $\alpha$ percent full load, $\beta$ percent reduced load, and $\varepsilon$ percent idle. Note that there are only two members in the numerator part of the equation—this is related to the fact that effective throughput during idle cycle is zero. It is important to remember that a reference test profile does not relate to any particular "use cycle" and is merely designed to highlight the elasticity of energy response. Nevertheless, the EER-VL metric (in Gbps/watt) gives a good idea of how the device will perform in the field under variable load conditions, even if the actual performance will vary.

Finally, it needs to be said that router and switch efficiency are perhaps most well-known but not the only equipment classes suitable for variable load metrics. In fact, just about any packet-oriented equipment (including firewalls, deep packet inspection devices, broadband, backhaul equipment, and base stations) can be characterized using a similar approach. It should be reasonably expected that over time, more equipment types will converge to a formula similar to (2) to capture their energy response with respect to load.

## VI. EXTENDED IDLE METRICS

While variable load metrics for telecom equipment are slowly converging towards usable implementation, a new class of green telecom capabilities is about to emerge. A foundation for this has been laid by the work within the IETF eman group [17] that formally defines energy management infrastructure for telecom devices, including that of explicit power states.

To understand this concept, it is important to remember that a variable load response to changing conditions is meant to happen in real time, which adds a lot of restrictions on what can and cannot be implemented to satisfy this requirement. For example, a whole group of power conservation methods based on depowering line cards and interfaces while proxying network states is not applicable to transit network equipment. Likewise, a popular concept of power-aware routing in the backbone network without using explicit power states is only capable of conserving energy within an elasticity window of bypassed network nodes (as shown in Table 2).

However, there are still many use cases where such restrictions can be partially lifted. A classic example is the enterprise environment, where distinct use cycles can be related to day/night or weekday/weekend shifts. Since such events are highly predictable, a telecom system can be programmed to safely degrade its performance, capacity, or quality of service during predicted low utilization periods. Should this programming be incorrect (imagine an odd working hour), the system will not be able to return to full capacity instantly, and service could be negatively impacted in the following transition period (Figure 4).

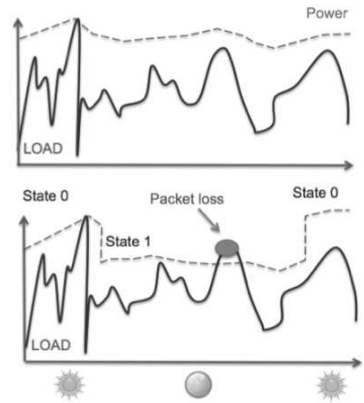

*Figure 4. Variable load (top) vs. extended idle (bottom) response*

As schematically depicted in Figure 4, a system transitioning through explicit power states has a chance to depower some slow moving components such as network processors, fabric planes, and memory banks. If the traffic unexpectedly surges, this may lead to packet loss.

It is easy to see that capabilities around managing explicit power states are orthogonal to that of variable load energy responses, and they in fact complement one another. Therefore, a new metric is required to capture this set of capabilities in an expression similar to (2) but modified for definition of power states. Moreover, if a telecom system is known to remain idle for an extended period of time, it should be switched off; hence, such a metric does not include a zero utilization period:

$$\text{EER-EX (Gbps/watt)} = \frac{(\alpha * T_f + \beta * T_{r1} + \varepsilon * T_{r2})}{(\alpha * E_{100} + \beta * E_{r1} + \varepsilon * E_{r2})} \quad (3)$$

where $T_f$ = full throughput (power state 0), $T_{r1}$ = reduced throughput in power state 1, $T_{r2}$ = reduced throughput in power state 2, and $\{E_{100}, E_{r1}, E_{r2}\}$ = energy consumption at full load, reduced load (state 1), and reduced load (state 2) respectively.

Although explicit power states are relatively new to large routers and switches, a very similar concept of capacity and speed degradation has actually been built into many telecom devices, from asymmetric digital subscriber line (ADSL+) power modes to energy management in cellular base stations. Therefore, we can reasonably expect this method of measuring efficiency to be extended over more device classes in future.

## VII. METRIC ROBUSTNESS

Up to this point, we have primarily focused on the scope and design of green metric formulas and their applicability to real-world equipment. In other words, we have attempted to develop a metric definition that is as clear and simple as possible without being too simplistic.

However, an example of TPR.9205 development highlights one important area that we have not touched on so far—namely, test methodology. A metric without a solid test methodology cannot be precise and scientifically accurate. In practical terms, this means that measurements of the same system at different test laboratories will not be repeatable, and measurements across different systems will not be comparable. Although a general metrology discussion is out of scope for this article, we should comment on one issue that every test methodology should try to avoid—loopholes. Loopholes are overlooked or unforeseen combinations of circumstances that might degrade the quality of measurement. Whether occurring on purpose or accidentally, methodology loopholes may threaten to invalidate the measurement set and skew the results.

Sometimes loopholes are easy to spot. For example, since a cooling system can draw a significant amount of power, a system under test that requires more cooling may produce worse efficiency results. A straightforward way to equalize test conditions for this would be to require certain environmental conditions such as ambient temperature and air pressure to be the same between test labs. However, it is not enough to solve the problem fully, as a system under test that was brought from a colder place may draw less energy until it warms up due to ambient air or because of internal heat sources. Such a simple difference can make or break the test.

A more complex example would be related to the distinction between variable load and extended idle energy management (as seen in Figure 4). For practical purposes, a test for variable load response of telecom equipment typically consists of three separate phases where utilization is well known. This choice is made simply for the ease of use. However, if the system under test is programmed to recognize and respond to periods of low utilization by changing its power states, it can "cheat" the test and get better results. Thus, a solid test method for assessing EER-VL would include the clause that obliges systems under test to be able to return to full capacity at any moment, with failure to do so invalidating the entire test.

In general, we should mention that robust test methodologies could only be developed over time and with the cooperation of parties interested in fair and unbiased comparisons. Failure to find a neutral test ground can also lead to failure in efficiency assessments.

## VIII. Conclusions and Future Work

In this paper, we have looked at the main developments in green telecom metrics. We have defined the need and purpose for peak, variable load, and extended idle metrics and described the evolution that some of those metrics have gone through.

This analysis lays a foundation for metric design in device categories not yet covered by well-defined energy efficiency measures. Understanding the needs and challenges of real-time versus non real-time operations is applicable to broad classes of telecom devices (one can easily see similarities between backhaul, mobile, and fixed wire line equipment). Likewise, the concept of explicit power states is already built into many product categories such as DSL modems and wireless base stations, and it only takes a proper formalization to describe the equipment in terms of energy response. Despite the fact that much work on energy assessment in telecom is still in its initial phase, we can conclude that progress is clearly being made and a sustainable ICT future is getting ever closer.